\documentclass[twocolumn,showpacs,superscriptaddress,prl,nofootinbib]{revtex4-1}

\usepackage[latin9]{inputenc}
\usepackage{amsmath}
\usepackage{amssymb}
\usepackage{amsthm}
\usepackage{graphicx}
\usepackage{color}
\usepackage{hyperref}
\hypersetup{%
    pdfborder = {0 0 0}
}

\theoremstyle{plain}
\newtheorem*{theorem*}{Theorem}

\makeatletter

\@ifundefined{textcolor}{}
{
 \definecolor{BLACK}{gray}{0}
 \definecolor{WHITE}{gray}{1}
 \definecolor{RED}{rgb}{1,0,0}
 \definecolor{GREEN}{rgb}{0,1,0}
 \definecolor{BLUE}{rgb}{0,0,1}
 \definecolor{CYAN}{cmyk}{1,0,0,0}
 \definecolor{MAGENTA}{cmyk}{0,1,0,0}
 \definecolor{YELLOW}{cmyk}{0,0,1,0}
}

\usepackage{color}
\usepackage{amsfonts}
\usepackage{graphics}
\usepackage{bm}\usepackage{bbm}
\usepackage{epsfig}\usepackage{amsthm}

\newcommand\blfootnote[1]{%
  \begingroup
  \renewcommand\thefootnote{}\footnote{#1}%
  \addtocounter{footnote}{-1}%
  \endgroup
}

\def\identity{\leavevmode\hbox{\small1\kern-3.8pt\normalsize1}}

\renewcommand{\epsilon}{\varepsilon}

\makeatother

\begin{document}
\title{Creating and concentrating quantum resource states in noisy environments using a quantum neural network}

\author{Tanjung Krisnanda}
\affiliation{School of Physical and Mathematical Sciences, Nanyang Technological University, 637371 Singapore, Singapore}
\blfootnote{Correspondence: T. Krisnanda (tanjungkrisnanda@gmail.com) or T. C. H. Liew (TimothyLiew@ntu.edu.sg)}

\author{Sanjib Ghosh}
\affiliation{School of Physical and Mathematical Sciences, Nanyang Technological University, 637371 Singapore, Singapore}
\author{Tomasz Paterek}
\affiliation{Institute of Theoretical Physics and Astrophysics, Faculty of Mathematics, Physics and Informatics, University of Gda\'{n}sk, 80-308 Gda\'{n}sk, Poland}
\author{Timothy C. H. Liew}
\affiliation{School of Physical and Mathematical Sciences, Nanyang Technological University, 637371 Singapore, Singapore}
\affiliation{MajuLab, International Joint Research Unit UMI 3654, CNRS, Universit\'{e} C\^{o}te d'Azur, Sorbonne Universit\'{e}, National University of Singapore, Nanyang Technological University, Singapore}

\begin{abstract}
Quantum information processing tasks require exotic quantum states as a prerequisite.
They are usually prepared with many different methods tailored to the specific resource state.
Here we provide a versatile unified state preparation scheme based on a driven quantum network composed of randomly-coupled fermionic nodes.
The output of such a system is then superposed with the help of linear mixing where weights and phases are trained in order to obtain desired output quantum states.
We explicitly show that our method is robust and can be utilized to create almost perfect maximally entangled, NOON, W, cluster, and discorded states.
Furthermore, the treatment includes energy decay in the system as well as dephasing and depolarization.
Under these noisy conditions we show that the target states are achieved with high fidelity by tuning controllable parameters and providing sufficient strength to the driving of the quantum network.
Finally, in very noisy systems, where noise is comparable to the driving strength, we show how to concentrate entanglement by mixing more states in a larger network.
\begin{flushleft}
\noindent {\bf Keywords}: Neural network applications, quantum neural network, optimization, quantum state preparation, quantum information, quantum entanglement, quantum machine learning.\\
\end{flushleft}

\end{abstract}

\maketitle

\noindent{\bf INTRODUCTION}

\noindent Quantum mechanics offers the possibility of performing tasks that are impossible in the classical regime.
This includes the pioneering works, to name a few, on quantum cryptography~\cite{cryptography}, quantum dense coding~\cite{dense-coding}, quantum teleportation~\cite{teleportation}, and quantum computing~\cite{jozsa2003role,nielsenchuang}.
Since then, a number of new proposals have been put forward, utilizing the resources available in the quantum regime, e.g., sensitive phase measurement in quantum metrology~\cite{resch2007time,slussarenko2017unconditional,nagata2007beating}, quantum memory \cite{fleischhauer2002quantum}, robust quantum computing~\cite{briegel2001persistent,Raussendorf2001,reimer2019high}, quantum secret sharing~\cite{hillery1999quantum}, or quantum computing without entanglement~\cite{datta2008quantum,gu2012observing,dakic2012quantum,pirandola2014quantum}.
All of the aforementioned proposals require pre-created quantum resource states.
Some exemplary ones include the maximally entangled states~\cite{RevModPhys.81.865}, NOON states~\cite{lee2002quantum}, W states~\cite{dur2000three}, cluster states~\cite{briegel2001persistent,Raussendorf2001,reimer2019high}, GHZ states~\cite{greenberger1989going}, and even discorded states~\cite{discord1,discord2}.
The supremacy of the proposed quantum solutions over what one can accomplish in the classical regime makes these proposals very attractive and motivates the creation of the corresponding quantum resource states.

Up to date, a variety of different methods were used to prepare such states: e.g., Refs.~\cite{wang2018multidimensional,kues2017chip,lu2020three} report high dimensional entanglement; NOON states were generated in  Refs.~\cite{walther2004broglie,mitchell2004super,afek2010high}; see also for W states~\cite{Weinfurter2004,grafe2014chip,fang2019three}; cluster states~\cite{walther2005experimental,kiesel2005experimental,mandel2003controlled,schwartz2016deterministic}; GHZ states~\cite{pan2000experimental,monz201114,kelly2015state}; and discorded states~\cite{dakic2012quantum,li2020deterministic}.
This motivates a unified approach in the creation of these resource states as it would provide a significant benefit over the diverse conventional methods.
Furthermore, these states possess quantum correlations, such as quantum entanglement, that are prone to noise.
This constitutes another challenge in their preparation, making it hard to achieve high fidelity.
Therefore, it is also important to analyze the creation of these states in noisy environments and, if their quality is limited, to enhance the quantum properties via schemes similar to entanglement distillation~\cite{bennett1996concentrating,bennett1996purification}.

On the other hand, biologically-inspired artificial neural networks excel at solving problems in the presence of noise and their flexibility allows their rapidly growing usefulness in many applications~\cite{webb2018deep,jones2019setting,topol2019high,rajpurkar2017cardiologist,nagy2019variational,hartmann2019neural,vicentini2019variational,mehta2019high}.
Among the different available artificial neural network architectures, reservoir computers are particularly striking in their ability to operate with randomly arranged nodes with no specific control of their coupling~\cite{Lukosevicius2012}. Instead, their functionality is attained by training a single set of connection weights corresponding to an output layer, which in practice corresponds to a simple linear processing of signals extracted from the network nodes. This method is less demanding as less control is required of the system, leading to several classical hardware demonstrations~\cite{tanaka2019recent,kusumoto2019experimental,ballarini2019polaritonic,rafayelyan2020large}.
Different forms of quantum neural networks have been developed for the preparation of quantum states using quantum computers~\cite{dash2020explicit}, and also applied in cryptography~\cite{jinjing2020approach}, while the general characteristics of quantum neural networks have also been explored~\cite{thecapacity}. The particular architecture of reservoir computing was also considered for use with quantum systems to perform quantum tasks such as the characterization of quantum states~\cite{qrpS,qrspsanjib,ghosh2020reconstructing,qnc}.

\begin{figure}[h]
\centering
\includegraphics[width=0.48\textwidth]{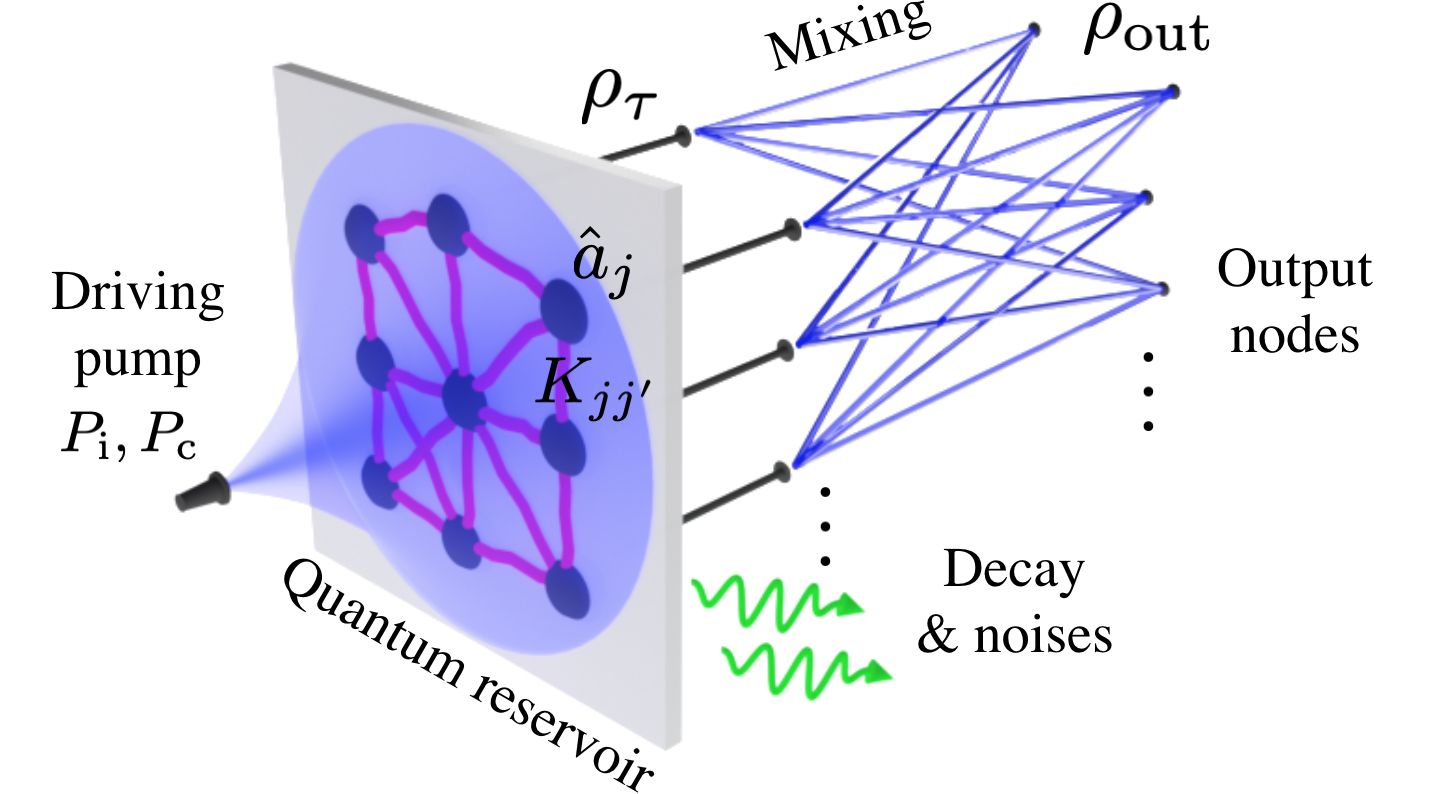}
\caption{Proposed general setup for the creation of quantum resource states.
A quantum reservoir network composed of fermions is driven either by coherent ($P_{\mbox c}$) or incoherent ($P_{\mbox i}$) pump.
The structure of the reservoir can be very flexible, i.e., with random individual energies and couplings ($K_{jj^{\prime}}$).
We also allow random interactions with environments, resulting in energy decay, dephasing and depolarizing of the reservoir state.
The signals from the reservoir then pass through a linear mixing setup.
The end product is output nodes containing the desired quantum state.
By training the mixing procedure, we show in the text that relevant quantum states, e.g., maximally entangled, NOON, W, cluster, GHZ, and discorded states, can be created and concentrated.
}
\label{FIG_setup}
\end{figure}

Motivated by this, here we propose the use of a quantum neural network as a unified platform to create and concentrate various quantum resource states in the presence of noise.
Our general scheme is depicted in Fig.~\ref{FIG_setup}.
It consists of a quantum network we shall refer to as the ``quantum reservoir'', which is made of a collection of fermionic nodes (e.g., quantum dots) that are interacting with random coupling strengths and are coherently or incoherently pumped.
We use the term reservoir for analogy with classical reservoir computing~\cite{Lukosevicius2012}, not to imply a large bath described by statistical averages.
We assume that the nodes emit a quantum signal, that is, they are quantum optical systems emitting light or quantum radiation.
As an example, quantum dots have been shown to emit quantum light (they are excellent single photon sources) \cite{claudon2010highly,he2013demand,somaschi2016near,ding2016demand,arita2017ultraclean,musial2020high}.
The signals can then be superposed, e.g., with a linear optical setup, giving rise to the desired quantum states as the final target output.
We note that quantum operations with linear optics (beam splitters and phase shifters) have been shown to be universal, i.e., being able to perform arbitrary unitary operations \cite{unitary,knill2001scheme,kok2007linear}, and have been demonstrated experimentally with tunable linear optical elements in Ref.~\cite{carolan2015universal}.
We also note an example of a stable integrated system where the signals from a quantum dot are demultiplexed and superposed using linear optics to successfully perform a task~\cite{wang2017high}.

We show that by training the weights and phases of the linear mixing with biologically-inspired algorithms, the output nodes contain high fidelity quantum resource states.
In most cases, our results show that noises coming from the interactions between the quantum reservoir and its environment are overcome when the driving strength exceeds the strength of the noise.
Furthermore, for weaker driving strength, we show a method to concentrate quantum entanglement of the imperfectly generated states. 
Remarkably, the reservoir does not need to be particularly large to engineer interesting and relevant quantum states.
Rather than define a specific system for implementation, we study a general but simple Hamiltonian, which could be realized with quantum dots, a range of cavity quantum electromagnetic systems, cold atoms, or superconducting qubit arrays.

\vspace{0.5cm}
\noindent{\bf PROPOSED SETUP}

\noindent Consider a quantum reservoir consisting of $N$ fermionic nodes arranged in a 2D setting with nearest neighbor interactions of random strength, see Fig.~\ref{FIG_setup}.
We note that one can also use bosonic nodes with strong nonlinear interactions, resulting in fermionized bosons~\cite{carusotto2009fermionized}.
Together with a coherent pump, the Hamiltonian describing our reservoir system, which will constitute the coherent part of its dynamics, can be written as
\begin{eqnarray}\label{EQ_rhamiltonian}
\hat H&=&\sum_j E_j \hat a_j^{\dagger} \hat a_j+\sum_{\langle jj^{\prime}\rangle} K_{jj^{\prime}} (\hat a_j^{\dagger} \hat a_{j^{\prime}}+\hat a_{j^{\prime}}^{\dagger} \hat a_{j}) \nonumber \\
&&+\sum_j (P_{\mbox{c},j} \hat a_j^{\dagger}+P_{\mbox{c},j}^* \hat a_j),
\end{eqnarray}
where $E_j$ and $K_{jj^{\prime}}$ are random local energies and nearest neighbor couplings of the fermionic systems, respectively. $|P_{c,j}|$ is the strength of the coherent driving.
We have used $\hat a_j$ ($\hat a_j^{\dagger}$) as the annihilation (creation) operator for the $j$th fermion.

Since our considered quantum reservoir is an open system the dynamics of its quantum state follows the Lindblad master equation
\begin{eqnarray}\label{EQ_Lmaster}
\dot \rho&\equiv&\mathcal{L}[\rho] \\
&=&-\frac{i}{\hbar}(\hat H\rho-\rho \hat H)+\sum_j \frac{\gamma_j}{\hbar} L[\rho,\hat a_j]+\frac{P_{\mbox{i},j}}{\hbar}L[\rho,\hat a_j^{\dagger}], \nonumber
\end{eqnarray}
where $\gamma_j$ and $P_{\mbox{i},j}$ are the energy decay and incoherent pumping strength for node $j$, respectively.
We have also used $L[\rho,\hat O]\equiv \hat O\rho \hat O^{\dagger}-\frac{1}{2}\hat O^{\dagger}\hat O \rho-\frac{1}{2}\rho \hat O^{\dagger}\hat O$.
As a result of interactions of the nodes with environment, in addition to energy decay, the nodes may also be dephased or depolarized (mixed with white noise) over time.
Taking these factors into account, the complete evolution of the state, in a coarse-grained equation, reads
\begin{equation}
\rho(t+\Delta t)= M_{\mbox{dp}}M_{\mbox{ds}}[\rho(t)+\Delta t \mathcal{L}[\rho(t)]],
\end{equation}
where $M_{\mbox{dp}}$ and $M_{\mbox{ds}}$ are depolarizing and dephasing channels, respectively, being applied in a continuous manner (see detailed expressions in Appendix A).
Similar to the energy decay term, for the $j$th node, we use $\gamma_{\mbox{ds},j}$ and $\gamma_{\mbox{dp},j}$ to denote the strength of the dephasing and depolarizing noise, respectively.
Note that all these noises affect the $j$th reservoir node locally, which is a natural assumption.

As the initial state of the system we assume no excitations, i.e., ground state for all individual nodes. 
We will show that, in some cases, this assumption is not necessary as we consider steady states, which are unique and independent of initial conditions.
In general, we let the system evolve from $t=0$ to $t=\tau$, at which we denote the state of the quantum reservoir as $\rho_{\tau}$.
The emission from each fermionic node is then split into different paths with the help of linear mixing, e.g., beam-splitters and phase shifters. 
After recombination of these paths, one ends up with output nodes, which contain the desired state $\rho_{\mbox{out}}$ (see Methods A for details of the linear mixing process).
The choice of the linear optical elements (weights and phases) is made using a training method that is described in Methods B. 
Essentially, we follow a biologically inspired genetic algorithm to maximize an output function. 
Here the natural choice for the output function is the fidelity of the prepared state with respect to the ideal state. 
To add more freedom, one can also perform a projective measurement on some unused output nodes (instead of simply tracing out) in order to obtain the final desired quantum state as a post-selected state.
We note that in our proposed setup, we assume the same polarization for the emitted photons. 
Although it would be interesting to generalize the formalism to include polarization as it would expand the available Hilbert space, the present scenario is already sufficient for the tasks that we considered.
Our treatment is similar to the ones already demonstrated experimentally, e.g., see Refs.~\cite{carolan2015universal,wang2017high}.

\vspace{0.5cm}
\noindent{\bf RESULTS}

\noindent{\bf A. Creating quantum resource states}

\noindent We now present the results on the creation of various resource states.
In order to quantify the quality of the output states produced we choose, as an output function, the fidelity between $\rho_{\mbox{out}}$ and an ideal target quantum resource state $\rho_{\mbox{tar}}$ (see Appendix B).
The fidelity gives a positive value, with a maximum of unity indicating equality between $\rho_{\mbox{out}}$ and $\rho_{\mbox{tar}}$.
In some cases, we will also directly maximize quantum correlations in the output state, such as quantum entanglement or quantum discord.

Moreover, note that the parameters involved for the quantum reservoir include the local node energies, the internode coupling strengths, and the strengths of the noises affecting each node.
We define a fixed quantity $\Gamma$ with energy unit, characterizing the quantum reservoir such that $E_j$, $K_{jj^{\prime}}$, $\gamma_j$, $\gamma_{\mbox{ds},j}$, and $\gamma_{\mbox{dp},j}$ are proportional to $\Gamma$ with the proportionality being a uniformly generated random number $\eta\in [0,1]$.
We also performed simulations with the random number sampled from an absolute value of a standard normal distribution (zero mean and standard deviation equals 1), in which case the conclusions of the paper remain unchanged.
In our simulations, independent $\eta$ is generated for all these different parameters and reservoir nodes.
This way, each parameter of the quantum reservoir is random with the maximum value given by $\Gamma$.
For simplicity, we also take constant and uniform driving pump, e.g., $P_{\mbox{c},j}$ or $P_{\mbox{i},j}= P$.
Our simulations generate different quantum resource states for different values of the controllable parameter $P$ relative to the energy unit $\Gamma$. 
The training of the linear mixing is performed with a hybrid random-genetic search algorithm, the details of which are given in Methods B.

\begin{figure}[!h]
\centering
\includegraphics[width=0.48\textwidth]{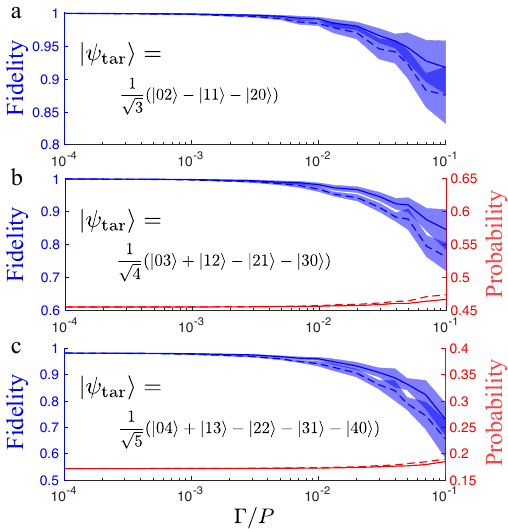}
\caption{Creation of maximally entangled states. We show the fidelity of the output states and the target ideal states $|\Psi_3^{\mbox{ME}}\rangle$ (a), $|\Psi_4^{\mbox{ME}}\rangle$ (b), and $|\Psi_5^{\mbox{ME}}\rangle$ (c) for different values of the ratio between the reservoir strength $\Gamma$ and the driving pump $P$. 
In all panels, the solid curves show the case with no dephasing and depolarizing noises,
while the dashed curves present the case where all the noises are taken into account.
The fidelity is averaged over 10 realizations of the reservoir parameters and the blue shaded regions are the corresponding standard deviations.
Each reservoir node is incoherently pumped with $P_{\mbox {i},j}=P$ and the time $\tau$ is chosen when the reservoir reaches steady state.
We consider a quantum reservoir having at most 4 fermionic nodes for all the panels.
For panels (b) and (c), the production of the corresponding entangled states is probabilistic, conditioned on having no excitations in the remaining output nodes.
The corresponding average probabilities are plotted as red curves.}
\label{FIG_maxe}
\end{figure}

\emph{\textbf{Maximally entangled states}}.---
We begin by considering entangled states. 
They are particularly useful in quantum cryptography~\cite{cryptography}, dense-coding~\cite{dense-coding}, teleportation~\cite{teleportation}, and also quantum computing~\cite{jozsa2003role,nielsenchuang}.
Higher dimensional quantum systems (qudits) provide higher capacity and more robustness against noise~\cite{cerf2002security,bouchard2017high,islam2017provably}. 
They also offer high efficiency and error tolerant quantum computing~\cite{lanyon2009simplifying}.
Here we consider quantum entanglement between two $N$-dimensional quantum objects $X$ and $Y$, the maximum of which is given by pure states of the form
\begin{equation}\label{EQ_maxe}
|\Psi^{\mbox{ME}}_N \rangle =\frac{1}{\sqrt{N}}\sum_k |X_k\rangle |Y_k\rangle,
\end{equation}
where $\{|X_k\rangle\}$ and $\{|Y_k\rangle\}$ form orthonormal bases for subsystems $X$ and $Y$, respectively.
Each component in the orthogonal superposition of Eq.~(\ref{EQ_maxe}) can have an arbitrary phase and this will not change quantum entanglement.
In our scenario of Fig.~\ref{FIG_setup} we consider bosonic emissions (photons) from the reservoir and since we are only interested in entanglement between two output nodes (principal nodes), we will measure the remaining nodes and post-select the output if the remaining nodes have zero photons (this way we maximize the number of photons in the principal nodes).
The creation of maximally entangled states using this method for up to $N=5$ is presented in Fig.~\ref{FIG_maxe}.
It is sufficient to consider 2, 3, and 4 randomly connected fermionic nodes for the quantum reservoir in order to target $|\Psi_3^{\mbox{ME}}\rangle$, $|\Psi_4^{\mbox{ME}}\rangle$, and $|\Psi_5^{\mbox{ME}}\rangle$, respectively.
The probability of post-selecting the target state is plotted by the red curves in Fig.~\ref{FIG_maxe} (b) and (c).

Let us discuss the effects of various sources of noise.
The solid blue curves in Fig.~\ref{FIG_maxe} are the results with energy decay only.
The dashed blue curves correspond to the situation where all the noises are present, causing further reduction in the fidelity.
Note that we plotted the average fidelity over 10 realizations of the random reservoir parameters and the blue shaded regions indicate the corresponding standard deviation.
Our results show that higher fidelities to the ideal states are obtained when the strength of the reservoir $\Gamma$ is weaker than the pump $P$.
This demonstrates that stronger pumping removes the effects causing imperfections.
Note also that the high fidelity regime is achieved by tuning the pump, which is a controllable parameter, making our method experimentally friendly.
Apart from fidelity, we also compute the corresponding value of entanglement for all the cases in Fig.~\ref{FIG_maxe} (cf. Appendix C).

\emph{\textbf{NOON states}}.---
NOON states were first terminologically used in quantum metrology by Lee \emph{et al.} \cite{lee2002quantum}.
This class of states allows one to perform supersensitive phase measurements, beyond the Heisenberg limit~\cite{resch2007time,slussarenko2017unconditional,nagata2007beating}.
Similar to the maximally entangled states, NOON states are defined for a bipartite system, and read
\begin{equation}
|\Psi^{\mbox{NOON}}_N \rangle=\frac{1}{\sqrt{2}}(|N0\rangle+|0N\rangle),
\end{equation}
up to an arbitrary relative phase, where $N$ is the number of excitations, e.g., photons.
These states are essentially superpositions of having $N$ photons in one mode or the other.
Therefore, they are also useful as $N$-photon sources \cite{munoz2014emitters}.
While the preparation of $|\Psi^{\mbox{NOON}}_2 \rangle$ can be done with the well-known Hong--Ou--Mandel setup, higher photon NOON states are harder to achieve.
Different methods have been used experimentally to prepare $N=3, 4$ \cite{walther2004broglie,mitchell2004super} and $N=5$ \cite{afek2010high} NOON states.

\begin{figure}[!h]
\centering
\includegraphics[width=0.48\textwidth]{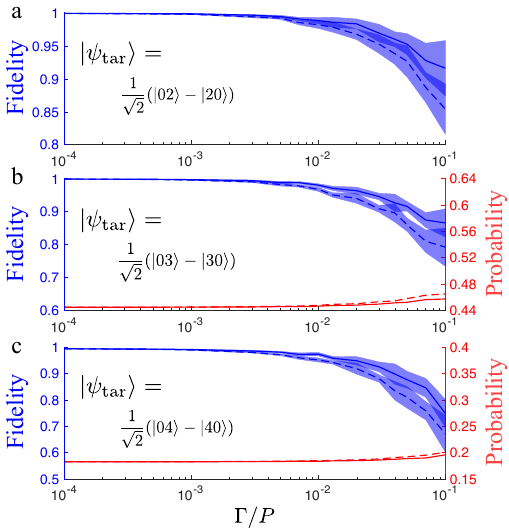}
\caption{Creation of NOON states. The fidelity is plotted for the target states indicated in the panels. Notation as in Fig.~\ref{FIG_maxe}.}
\label{FIG_noon}
\end{figure}

We present the generation of NOON states $|\Psi^{\mbox{NOON}}_N \rangle$ for up to $N=4$ in Fig.~\ref{FIG_noon}.
Similar to the entangled states, this arrangement requires at most 4 fermionic nodes for the quantum reservoir, all of which are incoherently pumped with $P_{\mbox{i},j}=P$.
We also note that $|\Psi^{\mbox{NOON}}_2 \rangle$ and $|\Psi^{\mbox{NOON}}_3 \rangle$ can be created with only 2 and 3 reservoir nodes.
The same trend is followed where high fidelity to the ideal states is obtained with high driving pump.
We note that $|\Psi^{\mbox{NOON}}_2 \rangle$ is obtained deterministically, while the generation is probabilistic for $|\Psi^{\mbox{NOON}}_3 \rangle$ and $|\Psi^{\mbox{NOON}}_4 \rangle$ with probabilities $\sim 0.44$ and $\sim 0.18$, respectively.

\emph{\textbf{W states}}.---
As an example of multipartite resource states we take the entangled multiple-qubit state known as the W state \cite{dur2000three}.
Due to their robust-to-loss property, these states are proposed for quantum storage~\cite{fleischhauer2002quantum}.
The $N$-qubit W states are written as
\begin{equation}
|\Psi^{\mbox{W}}_N\rangle=\frac{1}{\sqrt{N}}(|100\cdots 0\rangle+|010\cdots 0\rangle+\cdots+|00\cdots 01\rangle),
\end{equation}
where $N$ here is the number of qubits.

\begin{figure}[!h]
\centering
\includegraphics[width=0.48\textwidth]{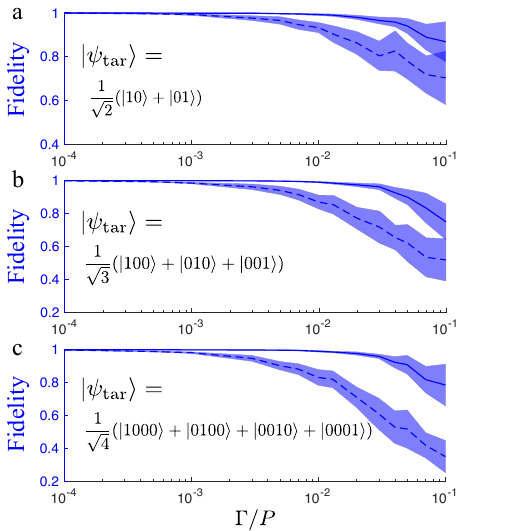}
\caption{Creation of W states $|\Psi^{\mbox{W}}_N\rangle$ for up to $N=4$. The 2-, 3-, and 4-qubit W states are presented in panel (a), (b), and (c), respectively. Here we simulate 20 sets of the random reservoir parameters. 
	All states are achieved deterministically. Notation as in Fig.~\ref{FIG_maxe}.}
\label{FIG_wstate}
\end{figure}

As our target states only have two orthogonal elements per node, i.e., either $|0\rangle$ or $|1\rangle$, here we consider fermionic output nodes.
With our method, we target $|\Psi^{\mbox{W}}_2\rangle$, $|\Psi^{\mbox{W}}_3\rangle$, and $|\Psi^{\mbox{W}}_4\rangle$, which are plotted in panel (a), (b), and (c) of Fig.~\ref{FIG_wstate}, respectively.
For all the panels, we only incoherently pump one reservoir node and wait until our quantum reservoir reaches a steady state.

\begin{figure}[!h]
\centering
\includegraphics[width=0.48\textwidth]{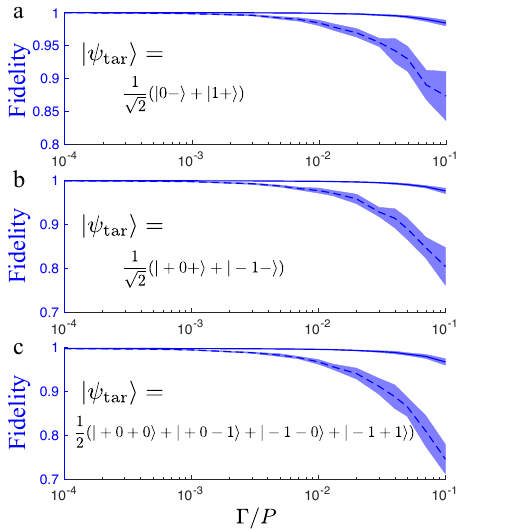}
\caption{Creation of cluster states.
Our simulations produce 20 different realizations of the reservoir parameters.
The average fidelity is plotted for $|\Psi^{\mbox{CL}}_2 \rangle$ (a), $|\Psi^{\mbox{CL}}_3 \rangle$ (b) and $|\Psi^{\mbox{CL}}_4 \rangle$ (c).
All states are obtained deterministically.
Notation as in Fig.~\ref{FIG_maxe}.}
\label{FIG_clusterstate}
\end{figure}

\emph{\textbf{Cluster states}}.---
Now we consider the so-called cluster states whose 2D versions are known to be universal for quantum computation~\cite{briegel2001persistent,Raussendorf2001}.
A few (2-, 3- and 4-qubit) cluster states are written as
\begin{eqnarray}
|\Psi^{\mbox{CL}}_2 \rangle&=&\frac{1}{\sqrt{2}}(|0-\rangle+|1+\rangle),\nonumber \\
|\Psi^{\mbox{CL}}_3 \rangle&=&\frac{1}{\sqrt{2}}(|+0+\rangle+|-1-\rangle),\nonumber \\
|\Psi^{\mbox{CL}}_4 \rangle&=&\frac{1}{2}(|+0+0\rangle+|+0-1\rangle+|-1-0\rangle \nonumber \\
&&+|-1+1\rangle),
\end{eqnarray}
where the subscript denotes the number of qubits and we have used $|\pm \rangle=(|0\rangle \pm|1\rangle)/\sqrt{2}$.
As these states are multi-qubit states, we will also consider fermionic output nodes in our method.
As the input, we consider coherent pumping of the quantum reservoir with pumping strength $P_{\mbox{c},j}=e^{i\pi/2}P/2$ for a duration $\tau=\pi \hbar/2P$, after which we proceed with the linear mixing.
We present the creation of $|\Psi^{\mbox{CL}}_2 \rangle$, $|\Psi^{\mbox{CL}}_3 \rangle$ and $|\Psi^{\mbox{CL}}_4 \rangle$ in panel (a), (b) and (c) of Fig.~\ref{FIG_clusterstate}, respectively.
All of the generated states are achieved deterministically, and we require a reservoir with at most 5 nodes.

At this stage we note that states that can be obtained via local unitary operations from the ones presented here are often easy to achieve. 
Along this line, we consider another important multi-qubit set of resource states known as the GHZ states.
GHZ states are good for secret communication between laboratories due to their sensitivity to loss \cite{hillery1999quantum}, a contrasting property to the W states.
For $N$ qubits, GHZ states read $|\Psi^{\mbox{GHZ}}_N \rangle = \frac{1}{\sqrt{2}}(|0\rangle^{\otimes N}+|1\rangle^{\otimes N})$, where $N\ge 3$ \cite{greenberger1989going}.
These states are fundamentally different from the W states in the sense that one cannot be created from another with local operations and classical communication \cite{dur2000three}.
However, a GHZ state of three qubits is related to the just-presented cluster state $|\Psi^{\mbox{CL}}_3\rangle$ via local unitaries and therefore can also be created with high fidelity.
A generic method for implementation of one-qubit gates in the framework of quantum reservoir processing has been recently proposed in Ref.~\cite{qnc}.

\newpage
\emph{\textbf{Discorded states}}.---
All of the target states mentioned so far were pure states.
Now we consider discorded states that are useful for certain quantum computing tasks \cite{datta2008quantum,gu2012observing,pirandola2014quantum}.
It has also been shown that discord is necessary for indirect creation of quantum entanglement between distant objects~\cite{dbound1, dbound2, revealing, distribution}.
Note that for pure states the concept of quantum discord reduces to quantum entanglement, and therefore the creation of maximum discord value follows the creation of maximally entangled states in Fig.~\ref{FIG_maxe}.
However, for general mixed states, these two quantum correlation properties can differ and it is of interest to generate discorded states without quantum entanglement.
This is because discorded states possess a quantum property that can persist in noisy environments, even when there is no quantum entanglement~\cite{discord1,discord2}.

\begin{figure}[!h]
\centering
\includegraphics[width=0.48\textwidth]{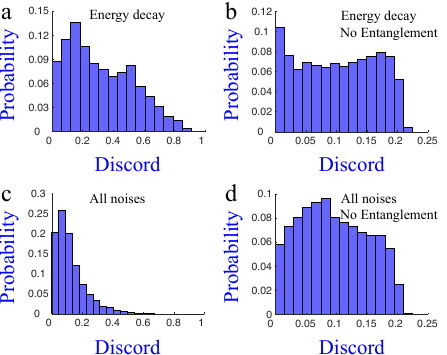}
\caption{Creation of discorded states. We consider a quantum reservoir having two nodes, one of which is incoherently pumped with $P_{\mbox{i},1}=P$.
We simulated $10^4$ different sets of the random reservoir parameters with $\Gamma = P$.
Panels (a) and (c) show the quantum discord between the output fermionic nodes without and with dephasing as well as depolarizing noises, respectively.
The corresponding quantum discord with an additional no-entanglement restriction is plotted in panels (b) and (d).
}
\label{FIG_discord}
\end{figure}

We present the creation of discorded states in Fig.~\ref{FIG_discord}, which shows the histogram of generated quantum discord over $10^4$ realizations of our reservoir parameters.
The quantum reservoir we consider consists of two fermionic nodes, where one of the nodes is incoherently pumped for a duration $\tau=\pi/P$.
In these simulations $P = \Gamma$, i.e., the pumping strength and noises in the system are comparable.
The linear mixing is then performed in order to maximize the relative entropy of discord (cf. Appendix B for discord quantifier) between the output fermionic nodes.
Panels (a) and (b) of Fig.~\ref{FIG_discord} show the creation of quantum discord for the case where only the energy decay is considered, while panels (c) and (d) are taking the additional dephasing as well as depolarizing noises into account.
In panels (b) and (d) we present the results under an additional constraint of no entanglement in the output state.

It has been shown previously that quantum discord is more robust against noise than quantum entanglement \cite{datta2008quantum,werlang2010thermal}.
Here, we note that for the case where the no-entanglement restriction is absent, i.e., panels (a) and (c) of Fig.~\ref{FIG_discord}, the quantum discord is greatly suppressed by the additional dephasing and depolarizing noises.
On the other hand, discorded states without the presence of entanglement do not feel the effect of these additional noises so severely, see panels (b) and (d).
In fact, when the only noise present is the energy decay (b), we see that lower values of discord are generated.
Our numerical results also show that maximum entangled states would not be attainable in the setup of Fig.~\ref{FIG_discord}, i.e., with the strength of noise being comparable to the driving pump of the quantum reservoir.
This motivates our next section.

\vspace{0.3cm}
\noindent{\bf B. Concentrating quantum resource states}

\noindent As we have shown, the ratio between the reservoir parameter $\Gamma$, which incorporates the strength of the noises, and the strength of the driving pump is an important indicator of high fidelity quantum resource states.
Here we propose a scheme to concentrate the imperfect states produced in the case where the parameter $\Gamma$ is comparable to the pump.
More generally, this scheme is useful in situations where the system parameters do not allow one to directly achieve the desired quantum state.
The property that we concentrate here is the quantum entanglement~\cite{RevModPhys.81.865}, which we quantify by negativity \cite{zyczkowski1998volume,lee2000entanglement,lee2000partial,negativity} (see~Appendix B).

\begin{figure}[h]
\centering
\includegraphics[width=0.48\textwidth]{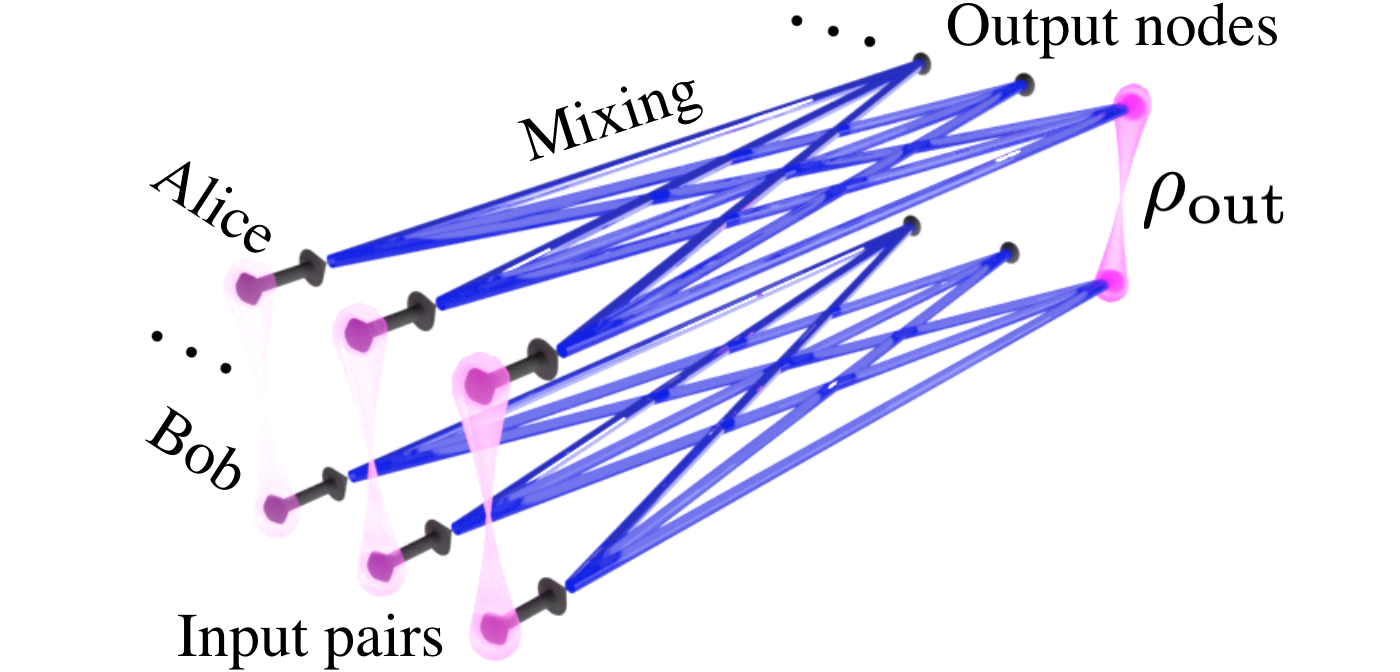}
\caption{Sketch of entanglement concentrating setup. We consider several imperfect pairs of entangled objects, e.g., the output of the generated states in Fig.~\ref{FIG_maxe} where the strength of the noise is comparable to the incoherent driving pump. One object of each pair is grouped together either in Alice's or Bob's side. In each side, we perform linear mixing where the coefficients are further trained to maximize quantum entanglement of an output pair.} 
\label{FIG_setupee}
\end{figure}

We take as input several pairs of imperfect states, 
for example those generated in Fig.~\ref{FIG_maxe}. 
Different particles of each pair are operated on by a different observer,
for example see Fig.~\ref{FIG_setupee} where the upper particles are operated by Alice and the lower particles by Bob.
Their aim is to 
achieve one pair of nodes with higher quantum entanglement than any of the input pairs.
For each party, the nodes are mixed, in particular, the coefficients and phases are trained, in order to maximize the entanglement of one pair in the output nodes, which is now represented by $\rho_{\mbox{out}}$.

\begin{figure}[!h]
\centering
\includegraphics[width=0.48\textwidth]{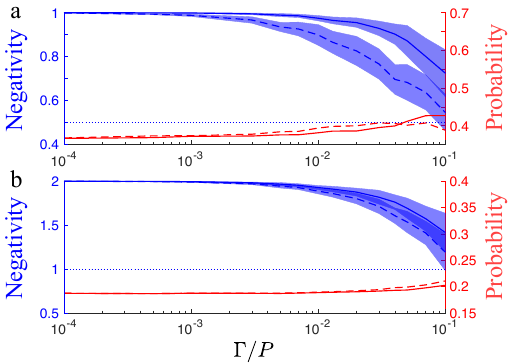}
\caption{Concentration of quantum entanglement.
(a) Final entanglement (as quantified by negativity) of the output nodes $\rho_{\mbox{out}}$ where only one reservoir node is incoherently pumped in the initial creation of entangled states.
(b) The corresponding results where two of the reservoir nodes are incoherently pumped.
Since we have two input pairs, all initial entanglement is concentrated onto the output pair in the limit $P \gg \Gamma$,
i.e., the concentrated entanglement is twice the input one in each pair.
Notation as in Fig.~\ref{FIG_maxe}.
The dotted lines in both panels show the maximum entanglement of the initial pair in the ideal case ($\Gamma/P=0$).}
\label{FIG_qree}
\end{figure}

In our simulations the input is taken as two copies of the entangled pair generated by the setup used to prepare Fig.~\ref{FIG_maxe}a.
The goal is to transfer the quantum entanglement to only one pair of output nodes, i.e., $\rho_{\mbox{out}}$ in Fig.~\ref{FIG_setupee}.
We present our entanglement concentrating results in Fig.~\ref{FIG_qree} for the case where one of the reservoir nodes is incoherently pumped (a) and where the pump is applied to two nodes (b).
Note that, as only two final output nodes are considered, we perform projective no-photon measurement on the rest of the nodes and plot the probability of successful post-selection as red curves.

The results show that our entanglement concentrating method can obtain as much quantum entanglement as possible even in the situation where $\Gamma/P\approx 0.1$.
We note that in the $\Gamma/P\rightarrow 0$ regime, the final maximum entanglement is double that of the maximum entanglement of each initial pair, hence perfect entanglement concentration is achieved.

\newpage
\vspace{0.5cm}
\noindent{\bf METHODS}

\noindent{\bf A. Linear mixing process}

\noindent Here we explain the process responsible for the mixing of the reservoir output $\rho_{\tau}$ and producing the output nodes, containing the state $\rho_{\mbox{out}}$, which is then compared with the desired target ideal state.
The properties calculated from the output density matrix $\rho_{\mbox{out}}$ include the fidelity to the ideal states, quantum entanglement, and quantum discord (explained in details in Appendix B).
Training of the mixing coefficients, in order to achieve the desired quantum resource states, will be explained separately in the next section.

The femionic nodes (e.g., quantum dots) in the quantum reservoir are two-level quantum objects, having annihilation (creation) operator denoted by $\hat a_j$ ($\hat a^{\dagger}_j$).
These nodes can emit, e.g., photons, which will be split and recombined to form the output nodes.
In this spirit, we note that linear mixing using linear optics (beam splitters and phase shifters) has been demonstrated experimentally in Ref.~\cite{carolan2015universal}.
We treat the emission from the reservoir nodes as bosons with the corresponding annihilation operator $\hat B_j$, satisfying the condition 
\begin{equation}\label{EQ_b1}
[\hat B_j,\hat B_{j^{\prime}}^{\dagger}]=\delta_{jj^{\prime}}.
\end{equation}
The output nodes after mixing are then represented by their annihilation bosonic operators $\hat C_k$, which are linear combinations of the operators $\hat B_j$, i.e.,
\begin{equation}\label{EQ_b2}
\hat C_k=\sum_j m_{kj}\: \hat B_j,
\end{equation}
where $m_{kj}$ are complex coefficients.
The coefficients $m_{kj}$ characterize the fractions (including additional phase shifters) of the reservoir emission nodes, forming the output nodes.

In particular, for $N$ independent reservoir nodes composing the quantum reservoir, one can create up to $N$ independent output nodes that must satisfy
\begin{equation}\label{EQ_b3}
[\hat C_k,\hat C_{k^{\prime}}^{\dagger}]=\delta_{kk^{\prime}}.
\end{equation}
By using Eqs. (\ref{EQ_b1}) and (\ref{EQ_b2}), one obtains, from Eq.~(\ref{EQ_b3}), the corresponding condition on the mixing coefficients:
\begin{equation}\label{EQ_matrixM}
\sum_j m_{kj} \:m^*_{k^{\prime}j}=\delta_{kk^{\prime}}.
\end{equation}
One can think of the coefficients $m_{kj}$ as elements of a matrix $M$.
Eq.~(\ref{EQ_matrixM}) gives the normalization and orthogonality of the rows (vectors) composing the matrix $M$.
These vectors form a complete basis since we consider the same number of output nodes as the reservoir nodes (i.e., a square matrix $M$).
One can show that the completeness of this basis ($\sum_u |u\rangle \langle u|=\mathbbm{1}$) results in $\sum_j m_{jk}^* \:m_{jk^{\prime}}=\delta_{kk^{\prime}}$, which together with Eq.~(\ref{EQ_matrixM}) implies $MM^{\dagger}=M^{\dagger}M=\mathbbm{1}$, and therefore $M$ is unitary.
As the linear mixing is a closed process, it preserves the excitation number, i.e.,
\begin{equation}
\mbox{tr}(\rho_{\tau}\sum_j \hat a_j^{\dagger} \hat a_j)=\mbox{tr}(\rho^{\mbox{bos}}_{\tau}\sum_j \hat B_j^{\dagger} \hat B_j)=\mbox{tr}(\rho_{\mbox{out}} \sum_k \hat C_k^{\dagger} \hat C_k),
\end{equation}
where $\rho^{\mbox{bos}}_{\tau}$ denotes the state of the reservoir $\rho_{\tau}$ embedded in the Hilbert space for the bosonic nodes.

The Fock states of the output nodes are computed as
\begin{equation}\label{EQ_nbasis}
|ab\cdots n\rangle=\frac{(C_1^{\dagger})^a (C_2^{\dagger})^b\cdots  (C_N^{\dagger})^n}{\sqrt{a! b! \cdots n!}}|\bm{0}\rangle,
\end{equation}
where $a$, $b$, and $n$ are the number of excitations of output nodes $1$, $2$, and $N$, respectively.
We have also used $|\bm{0}\rangle$ to denote the global vacuum state.
Given the reservoir state $\rho_{\tau}$ at time $\tau$, which is mapped to the state $\rho^{\mbox{bos}}_{\tau}$, one can compute the state of the output nodes $\rho_{\mbox{out}}$ in the new basis of Eq.~(\ref{EQ_nbasis}).
In particular, 
\begin{equation}\label{EQ_bb}
\rho_{\mbox{out}}^{ac\cdots n,a^{\prime}c^{\prime}\cdots n^{\prime}}=\langle ab\cdots n|\rho_{\tau}^{\mbox{bos}}|a^{\prime}c^{\prime}\cdots n^{\prime}\rangle.
\end{equation}
The resulting state $\rho_{\mbox{out}}$ in Eq.~(\ref{EQ_bb}) is implicitly dependent on the linear mixing coefficients $m_{kj}$ (elements of a unitary matrix $M$).
It should be apparent now that the linear mixing is a unitary (change of basis) excitation-preserving operation.
This allows one to strategize the input for the quantum reservoir in order to get the desired states, which possess the same number of excitations.
We note that this ``bosonic mixing" is also used in the entanglement concentrating setup.

The mixings for the W, cluster, and discorded states are done in a similar manner, with the exception that the operator for the output nodes $\hat C_k$ is now a fermionic operator, satisfying the anti-commutation condition $\{\hat C_k,\hat C_{k^{\prime}}^{\dagger}\}=\delta_{kk^{\prime}}$.
Following the steps described above, one can show that the corresponding mixing process is still unitary and preserving the excitation numbers.

\vspace{0.3cm}
\noindent{\bf B. Training}

\noindent The training in our scheme fixes the weights and phases (coefficients $m_{kj}$ of a matrix $M$) of the linear mixing, after the quantum reservoir state is evolved for a time $\tau$.
The purpose is to get the mixing coefficients $m_{kj}$ that maximize the desired output function, which is a property of $\rho_{\mbox{out}}$.
In our case, this is the fidelity to the ideal states, quantum discord or quantum entanglement.
In the former case, we maximize $\mathcal{F}=\langle \psi_{\mbox{tar}}|\rho_{\mbox{out}} |\psi_{\mbox{tar}} \rangle$ and in the latter $D_{1|2}$ or $E_{1:2}$ with the subscripts representing the considered output nodes (see~Appendix B for detailed expressions).

\begin{figure}[h]
\centering
\includegraphics[width=0.48\textwidth]{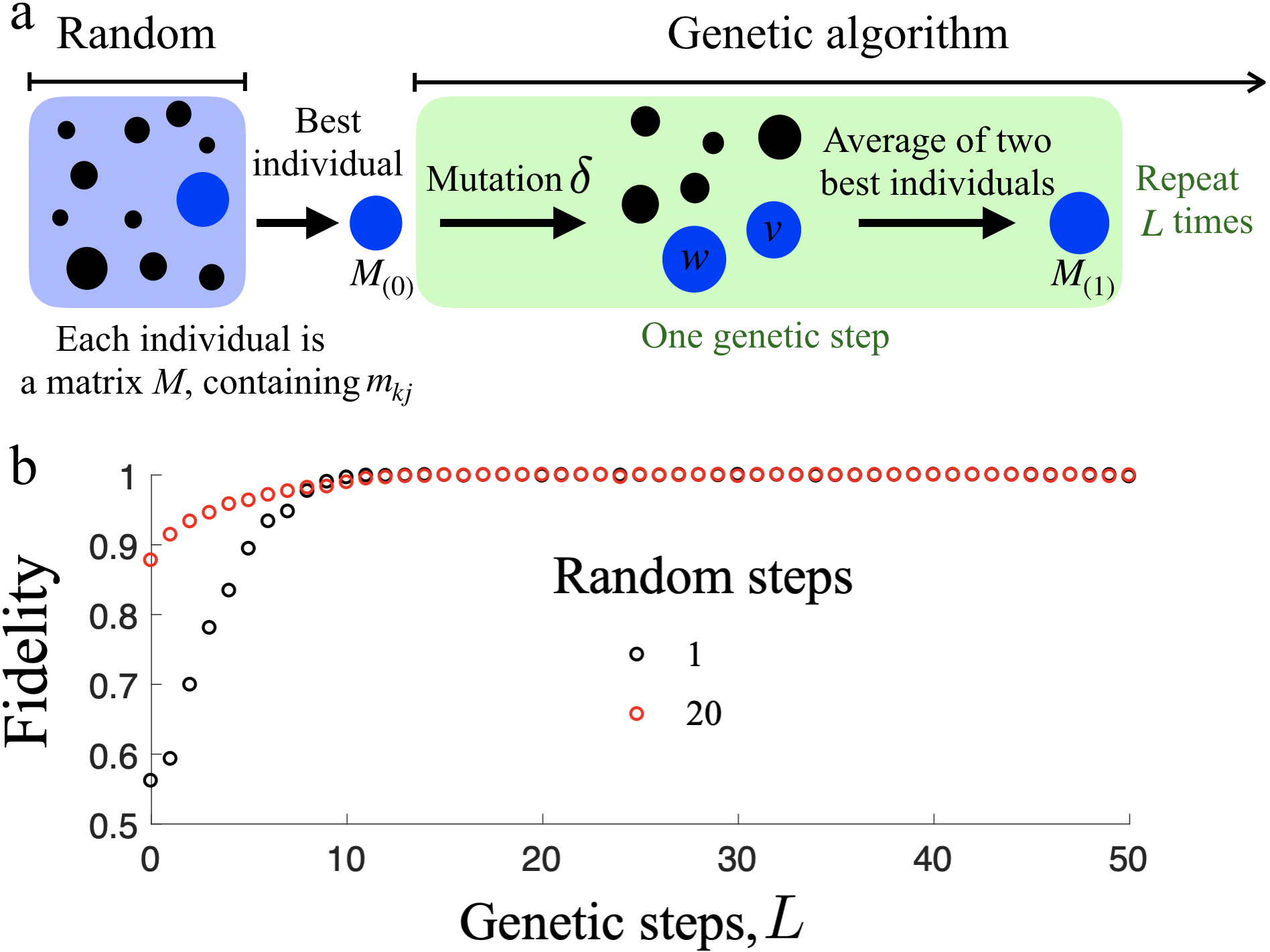}
\caption{\label{FIG_algo}(a) Illustration of random-genetic search algorithm to obtain the maximizing matrix $M$ containing the linear mixing coefficients $m_{kj}$. (b) Exemplary output function vs genetic steps with population size $S=10$ and mutation strength $\delta=0.01$.
Red (black) circles indicate the fidelity of creating the maximally entangled state (in the setup corresponding to panel (b) of Fig.~\ref{FIG_maxe}, in the limit $\Gamma/P\rightarrow 0$) for initial 20 (1) steps in the random algorithm.
Both cases can produce $\mathcal{F}>0.99$.}
\end{figure}

Our search for the maximizing parameters includes rough and smooth algorithms, as illustrated in Fig.~\ref{FIG_algo}a.
The rough algorithm is a random search (that involves generating random unitary matrices), producing good parameters, i.e., a matrix $M_{(0)}(\bm{\alpha}_0)$ (the first best individual in Fig.~\ref{FIG_algo}a), which will be used in the smooth search as the initial condition.
Here $\bm{\alpha}_0$ denotes the angular parameters for $M_{(0)}$.
As an example, a low dimensional $2\times 2$ matrix $M$ with its angular parameter $\bm{\alpha}=(\Theta,\Phi)$, can be parametrized as
\begin{equation}
M=\left(\begin{array}{cc}
\cos{\Theta}&\sin{\Theta}\:e^{i\Phi}\\
-\sin{\Theta}\:e^{-i\Phi}&\cos{\Theta}
\end{array}\right).
\end{equation}
We have used a subscript for $M$ to indicate the number of steps in the smooth algorithm.
We construct this search similar to the biologically inspired genetic algorithms, see Fig.~\ref{FIG_algo}a for illustration.
In particular, perturbations (mutations) around $M_{(0)}(\bm{\alpha_0})$ are added to produce new individuals,
\begin{equation}
P_{(L),i}= M_{(L-1)} (\bm{\alpha}_{L-1} +\delta {\bm \theta}_i),
\end{equation}
where $P_{(L),i}$ denotes a perturbed matrix in the $L$th step with random angles ${\bm \theta}_i$ and a small number $\delta$ characterizing the strength of mutation.
The index $i=1,2,\cdots S$ enumerates individuals in new sets with population size $S$.
We then compute the matrix $M_{(L)}$ from the average of the two individuals (labeled, $v$ and $w$) having the highest and second highest output function, respectively, i.e.,
\begin{equation}
M_{(L)}=M_{(L-1)}(\tilde {\bm{\alpha}}_{L-1}),
\end{equation}
with the angle
\begin{equation}
\tilde {\bm{\alpha}}_{L-1}= {\bm{\alpha}}_{L-1}+\delta\:\frac{{\bm \theta}_v+{\bm \theta}_w}{2}.
\end{equation}

As an example, we present how our algorithm works for the creation of maximally entangled states.
In Fig.~\ref{FIG_algo}b we illustrate how the fidelity to the ideal state $|\Psi^{\mbox{ME}}_4\rangle$ (as in the setup for Fig.~\ref{FIG_maxe}b) is converging to a maximum value ($\rightarrow 1$) as we increase the number of steps $L$.
We consider the population size of $S=10$ for the genetic search.
It is apparent that increasing the steps for the initial rough search increases the initial value for the output function.
However, higher fidelity, e.g., $\mathcal{F}>0.99$ is hard to achieve with only random search with $500$ steps.

\vspace{0.5cm}
\noindent{\bf DISCUSSION}

\noindent{\bf A. Related works}

\noindent The proposals on quantum cryptography~\cite{cryptography}, dense-coding~\cite{dense-coding}, teleportation~\cite{teleportation}, and quantum computing~\cite{jozsa2003role,nielsenchuang} have made the preparation of entangled states a crucial task.
Moreover, the use of higher dimensional quantum systems has attracted attention as they possess higher capacity and are less prone to noise, as shown in Refs.~\cite{cerf2002security,bouchard2017high,islam2017provably,lanyon2009simplifying}. 
This has motivated the creation of bipartite high dimensional entangled states, as those written in Eq.~(\ref{EQ_maxe}) or their equivalent, e.g., see Refs.~\cite{wang2018multidimensional,kues2017chip,lu2020three}.
In particular, for $N=3,4,$ and $5$, the entangled states were prepared with fidelities $0.98, 0.96,$ and $\approx 0.94$, respectively~\cite{wang2018multidimensional}.
In Fig.~\ref{FIG_maxe}, in the low $\Gamma/P$ regime, our quantum network generated the entangled states for $N=3$ and $4$ with fidelity $>0.999$, while $N=5$ with $0.982$.
We note that our results are from theoretical simulations.

NOON states allow phase measurement in quantum metrology with super sensitivity and super resolution~\cite{resch2007time,slussarenko2017unconditional,nagata2007beating}.
The generation of NOON states have been reported in Refs.~\cite{walther2004broglie,mitchell2004super,afek2010high}.
For example, the ideal fidelities proposed in Ref.~\cite{afek2010high} are $1,1$, and $0.933$ for the preparation of NOON states with $N=2,3$, and $4$, respectively.
Our simulations (see Fig.~\ref{FIG_noon}) show the corresponding fidelities can reach $0.999$ for $N=2$ and $3$, and $0.994$ for $N=4$.

The scheme for quantum memory requires robust-to-noise W states~\cite{fleischhauer2002quantum}. 
Reports on the state preparation can be seen in Refs.~\cite{Weinfurter2004,grafe2014chip,fang2019three}.
In particular, probabilistic generation of W states for up to $16$ qubits has been reported in Ref.~\cite{grafe2014chip}.
Recently, a preparation of three-qubit W state has been achieved with a fidelity of 0.92~\cite{fang2019three}.
In our case, as can be seen in Fig.~\ref{FIG_wstate}, the 2-, 3-, and 4-qubit W states were achieved deterministically with fidelity $>0.997$ in the low $\Gamma/P$ limit.

The pursuit of robust quantum computing has motivated the introduction of cluster states~\cite{briegel2001persistent,Raussendorf2001,reimer2019high}.
These states have been generated in Refs.~\cite{walther2005experimental,kiesel2005experimental,mandel2003controlled,schwartz2016deterministic}.
For example, the preparation of 3- and 4-qubit cluster states was demonstrated with fidelity $0.69$~\cite{schwartz2016deterministic} and $0.74$~\cite{kiesel2005experimental}, respectively.
In Fig.~\ref{FIG_clusterstate}, the cluster states can be achieved with fidelity $>0.999$ for 2- and 3-qubit cases, and $0.997$ for 4-qubit cluster state.

For quantum computing without entanglement, discorded states have been proposed as a resource~\cite{datta2008quantum,gu2012observing,pirandola2014quantum}.
These states are normally mixed states that are more robust to noise, as compared to entangled states.
See, for example Refs.~\cite{dakic2012quantum,li2020deterministic}, for preparation of states with quantum discord. 
In this paper, we did not specifically target particular states, rather we showed a distribution of discorded states produced by our method in Fig.~\ref{FIG_discord}.

\vspace{0.3cm}
\noindent{\bf B. Physical implementations}

\noindent Let us now discuss possible physical implementations of our method.
The reservoir network consisting of fermionic nodes, e.g., following the Hamiltonian in Eq.~(\ref{EQ_rhamiltonian}), can be realized using an interacting quantum dot system~\cite{bimberg1999quantum}, atoms trapped in an optical lattice~\cite{esslinger2010fermi,hofstetter2018quantum,tarruell2018quantum}, fermionized photons in an array of nonlinear cavities~\cite{carusotto2009fermionized,bardyn2012majorana}, Rydberg atoms in a network of connected cavities~\cite{chang2014quantum}, interacting polaritonic~\cite{photoniccrystal,delteil2019towards} or trion polaritonic systems~\cite{emmanuele2019highly,kyriienko2019nonlinear}, microwave photons in coupled superconducting resonators \cite{vaneph2018observation}, or arrays of atoms~\cite{poshakinskiy2020quantum}.


One can also couple these potential quantum reservoir systems to waveguides or resonators, see for example Refs.~\cite{javadi2015single,katsumi2018transfer,schnauber2019indistinguishable,turschmann2019coherent,roy2017colloquium}.
The arbitrary linear mixing can be done with linear optics, i.e., beam-splitters and phase shifters~\cite{unitary,knill2001scheme,kok2007linear} with tunable elements~\cite{carolan2015universal} or through tunable coupled waveguides~\cite{bishop2018electro}, which provide a versatile and scalable platform.
For quantum dots, we note previous experimental demonstration with stable and well integrated linear optics elements to perform multiphoton boson sampling~\cite{wang2017high}.
See also Ref.~\cite{wang2019integrated} for a comprehensive review on integrated photonic systems for quantum technologies.
In the case of fermionic output, the linear mixing can also be realized via coupling between the quantum reservoir and the output nodes, e.g., coupled fermionized cavities \cite{carusotto2009fermionized,bardyn2012majorana}.

\vspace{0.5cm}
\noindent{\bf SUMMARY}

\noindent We have shown that a variety of quantum resource states can be created with a quantum reservoir neural network with suitable pump and training of a linear mixer.
We take into account the randomness of the reservoir parameters, including the strength of the noises and show that high quality states are produced in the limit where the pumping power applied on the quantum reservoir is stronger than the imperfections coming out from the noises.
In the case where noises are inevitable and significant, i.e., their amplitudes are comparable to the pumping strength, imperfect states would be generated.
We show that by having more copies of these states and performing local mixing, one can refocus the quantum entanglement to only one pair of output nodes.

We note that our method does not require precise control over the parameters of the quantum reservoir, which includes the local node energies, couplings between nodes, and the noises.
It is sufficient to only have control over the input, i.e., the coherent or incoherent pumping, and the linear mixing of the output layer, making our method experimentally friendly.
Along the direction of our proposed method, there are many more possibilities of creating other states or manipulating the generated states.
For example, after the first linear mixing in Fig.~\ref{FIG_setup}, one can continue further mixing to do state manipulation.
In principle, our setup can also be integrated in order to perform specific quantum tasks.

\vspace{0.2cm}
\noindent{\bf ACKNOWLEDGEMENTS}

\noindent We would like to thank Daniele Sanvitto and Micha{\l}  Matuszewski for stimulating discussions.
T.K., S.G., and T.C.H.L were supported by the Singapore Ministry of Education Academic Research Fund Project No. MOE2017-T2-1-001. T.P. was supported by the Polish National Agency for Academic Exchange NAWA 470 Project No. PPN/PPO/2018/1/00007/U/00001.

\vspace{0.1cm}
\noindent{\bf Author contributions:} T.K., S.G., and T.C.H.L. conceived the initial project direction; T.K. carried out the calculations and derivations under supervision of S.G. and T.C.H.L.; T.K. wrote the paper with the help from S.G., T.P., and T.C.H.L.; All authors discussed and analyzed the results.

\vspace{0.1cm}
\noindent{\bf Competing interests} 

The authors declare no competing financial and non-financial interests.

\vspace{0.1cm}
\noindent {\bf Data availability} 

All data needed to evaluate the conclusions in the paper are present in the main text.

\vspace{0.1cm}
\noindent {\bf Code availability} 

Codes are available upon request from the authors.

\vspace{0.5cm}
\noindent{\bf APPENDIX}

\noindent{\bf A. Noises from interactions with environments}

\noindent As the target quantum states contain quantum correlations that are prone to decay and noises, it is important to take these imperfections into account in our proposed setup.
Here we describe the noises that may affect the quantum reservoir (see Ref.~\cite{nielsenchuang} for a review), resulting from interactions between the reservoir nodes and their environments.
This includes the dephasing and depolarization of the reservoir state, which corresponds to loss of quantum information and mixing the state with white noise, respectively.
Additionally, the reservoir state may also lose energy, which is often described by an amplitude damping process.
In our scenario, this effect is already incorporated in the decay term $\gamma_j$ in Eq.~(\ref{EQ_Lmaster}).

\emph{\textbf{Dephasing noise}}.---
The loss of coherence is seen from the decreasing strength of the off-diagonal terms in the density matrix describing the reservoir state.
For a state $\rho(t)$, the dephasing process for an infinitesimal time $\Delta t$ is described with the following channel:
\begin{equation}
\rho(t+\Delta t)=M_{\mbox{ds}}[\rho(t)]=M_{\mbox{ds},N}\cdots M_{\mbox{ds},2}M_{\mbox{ds},1}[\rho(t)],
\end{equation}
where the application of each channel reads
\begin{equation}
M_{\mbox{ds},j}[\rho(t)]=\left(1-\frac{\gamma_{\mbox{ds},j}}{2\hbar}\Delta t\right)\rho(t)+\frac{\gamma_{\mbox{ds},j}}{2\hbar} \Delta t \:\sigma_{z,j}\rho(t) \:\sigma_{z,j}.
\end{equation}
We have used $\gamma_{\mbox{ds},j}$ and $\sigma_{z,j}$ to denote the dephasing strength and $z$-Pauli matrix for the $j$th reservoir node.

\emph{\textbf{Depolarizing noise}}.---
The depolarizing process is modelled by mixing the reservoir state with the local maximally mixed state, i.e., local white noise.
In our case, the application of this channel to state $\rho(t)$ reads
\begin{equation}
\rho(t+\Delta t)=M_{\mbox{dp}}[\rho(t)]=M_{\mbox{dp},N}\cdots M_{\mbox{dp},2}M_{\mbox{dp},1}[\rho(t)],
\end{equation}
where we use
\begin{eqnarray}
M_{\mbox{dp},j}[\rho(t)]&=&(1-\frac{\gamma_{\mbox{dp},j}}{\hbar}\Delta t)\rho(t) \nonumber \\
&&+\frac{\gamma_{\mbox{dp},j}}{3\hbar} \Delta t \:\left(\sum_{\alpha=x,y,z}\sigma_{\alpha,j}\rho(t) \:\sigma_{\alpha,j}\right),
\end{eqnarray}
with $\gamma_{\mbox{dp},j}$ denoting the strength of the depolarizing process for reservoir node $j$ and $\sigma_{\alpha,j}$ is $\alpha$-Pauli matrix.

\vspace{0.3cm}
\noindent{\bf B. Fidelity and quantum correlations}

\emph{\textbf{Fidelity}}.---
\noindent Quantum fidelity is used as a measure to quantify the success probability for a quantum state $\rho$ to be identified as another state $\sigma$.
We use the square fidelity~\cite{jozsa1994fidelity}:
\begin{equation}
\mathcal{F}(\rho,\sigma) \equiv \left(\mbox{tr}\left( \sqrt{\sqrt{\sigma}\rho\sqrt{\sigma}}\right)\right)^2.
\end{equation}
This is a symmetrical quantity, i.e., $\mathcal{F}(\rho,\sigma)=\mathcal{F}(\sigma,\rho)$, and its value ranges from zero, for two distinguishable states, to unity, achieved if and only if $\rho=\sigma$.

In most cases presented in the main text, fidelity is computed between a (in general) mixed quantum state $\rho_{\mbox{out}}$ and a pure target state $\rho_{\mbox{tar}}=|\psi_{\mbox{tar}}\rangle \langle \psi_{\mbox{tar}}|$.
In this case the expression of fidelity reduces to $\mathcal{F}=\langle \psi_{\mbox{tar}}|\rho_{\mbox{out}} |\psi_{\mbox{tar}} \rangle$.

\emph{\textbf{Quantum entanglement}}.---
A bipartite system is entangled if its state cannot be described by a separable form (cf. Ref.~\cite{RevModPhys.81.865} for a review)
\begin{equation}
\rho_{\mbox{sep}}=\sum_j p_j\: \rho_{X,j}\otimes \rho_{Y,j},
\end{equation}
where $\rho_{X,j}$ ($\rho_{Y,j}$) is a state for system $X$ ($Y$) and $\{p_j\}$ form a probability distribution.

In order to quantify the entanglement we use negativity \cite{zyczkowski1998volume,lee2000entanglement,lee2000partial,negativity}, which is defined as
\begin{equation}
E_{X:Y}(\rho)=\frac{||\rho^{T_Y}||_1-1}{2},
\end{equation}
where $T_Y$ denotes partial transposition with respect to $Y$ and $||\cdot||_1$ is a trace norm.

\emph{\textbf{Quantum discord}}.---
Quantum discord is defined to be present in states which are not so-called quantum-classical between systems $X$ and $Y$, i.e., not in the form (cf. Refs.~\cite{discord1,discord2} for reviews):
\begin{equation}\label{EQ_qcstates}
\rho_{\mbox{qc}}=\sum p_j\: \rho_{X,j}\otimes |j\rangle_Y\langle j|,
\end{equation}
where $\{|j\rangle_Y\}$ form an orthonormal basis for system $Y$.
States that are not of the above class admit discord $D_{X|Y}>0$.
Note that unlike negativity, this discord measure is not symmetrical under the exchange of $X$ and $Y$.

In the main text we use the relative entropy of discord as its quantifier~\cite{modi2010unified,deficit}, which reads
\begin{equation}
D_{X|Y}(\rho)=\min_{\rho_{\mbox{qc}}}[ -\mbox{tr}(\rho \log(\rho_{\mbox{qc}}))] +\mbox{tr}(\rho \log(\rho)),
\end{equation}
where the minimization is over the quantum-classical states of Eq.~(\ref{EQ_qcstates}).

\vspace{0.3cm}
\noindent{\bf C. Quantum entanglement of the output states}

\noindent Here we provide the calculations of quantum entanglement as quantified by negativity for the scheme of creating maximally entangled states in Fig.~\ref{FIG_maxe}. This illustrates how another figure of merit behaves in our setup.

\begin{figure}[!h]
\centering
	\includegraphics[width=0.45\textwidth]{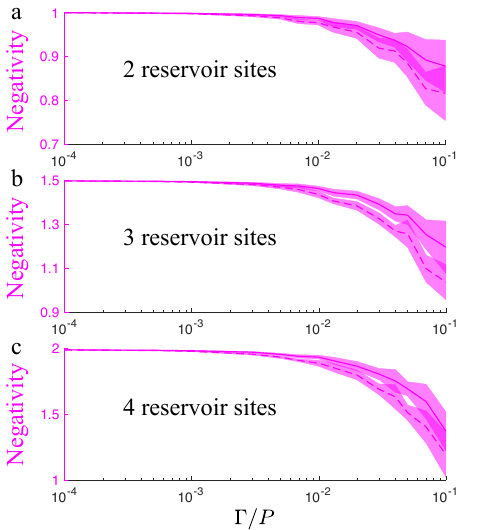}
	\caption{Quantum entanglement as quantified by negativity for the schemes in Fig.~\ref{FIG_maxe}. Panels (a), (b), and (c) correspond to 2, 3, and 4 reservoir nodes being incoherently pumped. Notation as in Fig.~\ref{FIG_maxe}.}
	\label{FIG_maxeneg}
\end{figure}

We present quantum entanglement of the generated output states $\rho_{\mbox{out}}$ in Fig.~\ref{FIG_maxeneg} where 2 (a), 3 (b), and 4 (c) of the reservoir fermionic nodes are pumped incoherently.
Note that in the ideal case, i.e., $\Gamma/P\rightarrow0$, the maximum entanglement is close to $1$, $1.5$, and $2$ in panels (a), (b), and (c), respectively.
These are the corresponding quantum entanglements of the ideal target states $|\Psi^{\mbox{ME}}_3\rangle$, $|\Psi^{\mbox{ME}}_4\rangle$, and $|\Psi^{\mbox{ME}}_5\rangle$, respectively.

\end{document}